\documentclass[a4paper,11pt]{article}
\usepackage{pos}
\usepackage{hhline}

\newcommand{\mukin}{\mu_\pi^2}
\newcommand{\muG}{\mu_G^2}
\newcommand{\darwin}{\rho_D^3}
\newcommand{\Opsix}[2]{O_{#1}^{#2}}

\newcommand{\OpsevenP}[2]{P_{#1}^{#2}}

\newcommand{\intm}{\textrm{int}^{-}}
\newcommand{\intp}{\textrm{int}^{+}}
\newcommand{\exc}{\textrm{exc}}
\newcommand{\GeV}{\,\textrm{GeV}}

\title{Charmed hadron lifetimes}

\author[a]{James Gratrex}
\author*[a]{Bla\v zenka Meli\' c}
\author[a]{Ivan  Ni\v sand\v zi\'c}

\affiliation[a]{Ru\dj er Bo\v skovi\'c Institute, Bijeni\v cka cesta 54, 10000, Zagreb, Croatia.}

\emailAdd{jgratrex@irb.hr}
\emailAdd{blazenka.melic@irb.hr}
\emailAdd{ivan.nisandzic@irb.hr}

\abstract{We provide updated predictions of the lifetimes of singly charmed baryons and mesons within the heavy quark expansion, with all known corrections included. A special attention is devoted to the choice of the charm mass and wavefunctions of heavy baryons.
Our results accommodate the experimentally-favoured hierarchy of singly charmed baryon lifetimes 
\begin{eqnarray*}
\tau(\Xi_c^0) < \tau(\Lambda_c^+)< \tau(\Omega_c^0) < \tau(\Xi_c^+)\,
\end{eqnarray*}
in contrast to earlier theoretical findings. 
Predictions for charmed meson lifetimes and semileptonic branching ratios are also in agreement, within uncertainties, with a recent comprehensive study and with experimental results.}

\FullConference{%
  Corfu Summer Institute 2022 "School and Workshops on Elementary Particle Physics and Gravity",\\
  28 August - 1 October, 2022\\
  Corfu, Greece}

\begin{document}
\maketitle

\section{Introduction}
Lifetimes of weakly-decaying particles containing a heavy (charm or bottom) quark have long played a significant role in driving the development of experimental and theoretical particle physics, including several of the earliest `anomalies' observed in the flavour sector. For example, in the late 1970s, the ratio of lifetimes of the $D^+$ and $D^0$ mesons was measured to differ significantly from the naive expectation of one. Using modern values, it reads \cite{PDG2022}
\begin{equation}
    \frac{\tau(D^+)}{\tau(D^0)} = 2.54(2) \, .
\end{equation}
This deviation from the initial expectation was explained in the early 1980s as arising from a large Pauli interference contribution to the $D^+$ lifetime \cite{GNPR1979,KS1983,BGT1984,SV1985,SV1986}, eventually leading to the development of the heavy quark expansion (HQE), see e.g. \cite{Lenz2014} for a review. Furthermore, in the mid-1990s, within the HQE, it was found that the lifetimes of $b$-hadrons should not differ by more than 10\%, so it was a surprise that the experimental ratio of the $\tau(\Lambda_b^0)/\tau(B_d^0)$ was measured to be approximately 0.75(5). Again, theory won the day, as later measurements have seen this ratio return to the theoretical expectation \cite{SV1986,GLMNPR2023}.

Recently, another potential anomaly in lifetimes has appeared. LHCb measurements of charmed baryon lifetimes \cite{LHCbOmegac2018,LHCbcharmedLifetimes2019,LHCbOmegac2021} indicate that the $\Omega_c^0$ lifetime is four times larger than, and wholly inconsistent with, the earlier experimental results \cite{FOCUSOmegac2003,SELEXOmegac}. Since existing theoretical predictions \cite{SV1986,GRT1986,Melic97c,Cheng97c,Cheng18c} tended to support a shorter lifetime for the $\Omega_c^0$, this motivates a reassessment of the prediction. 

In this proceeding, we present new predictions of inclusive observables for baryons containing a single charm quark, based on the recent work \cite{GMN2022}. Compared with the previous predictions in \cite{SV1986,GRT1986,Melic97c,Cheng97c,Cheng18c}, we include newly-available contributions, such as the Darwin contribution, recently made available for nonleptonic and charmed decays in \cite{LPR2020,MMP2020,Moreno2020,KLPRRV2021}, $1/m_c$ corrections to the four-quark operators, and $\alpha_s$ corrections to the Wilson coefficients of two- and four-quark operators. We also apply the same calculation to mesons, with results agreeing with the recent work \cite{KLPRRV2021}.

\begin{figure}[htb]
    \centering
    \includegraphics[width=0.8\textwidth]{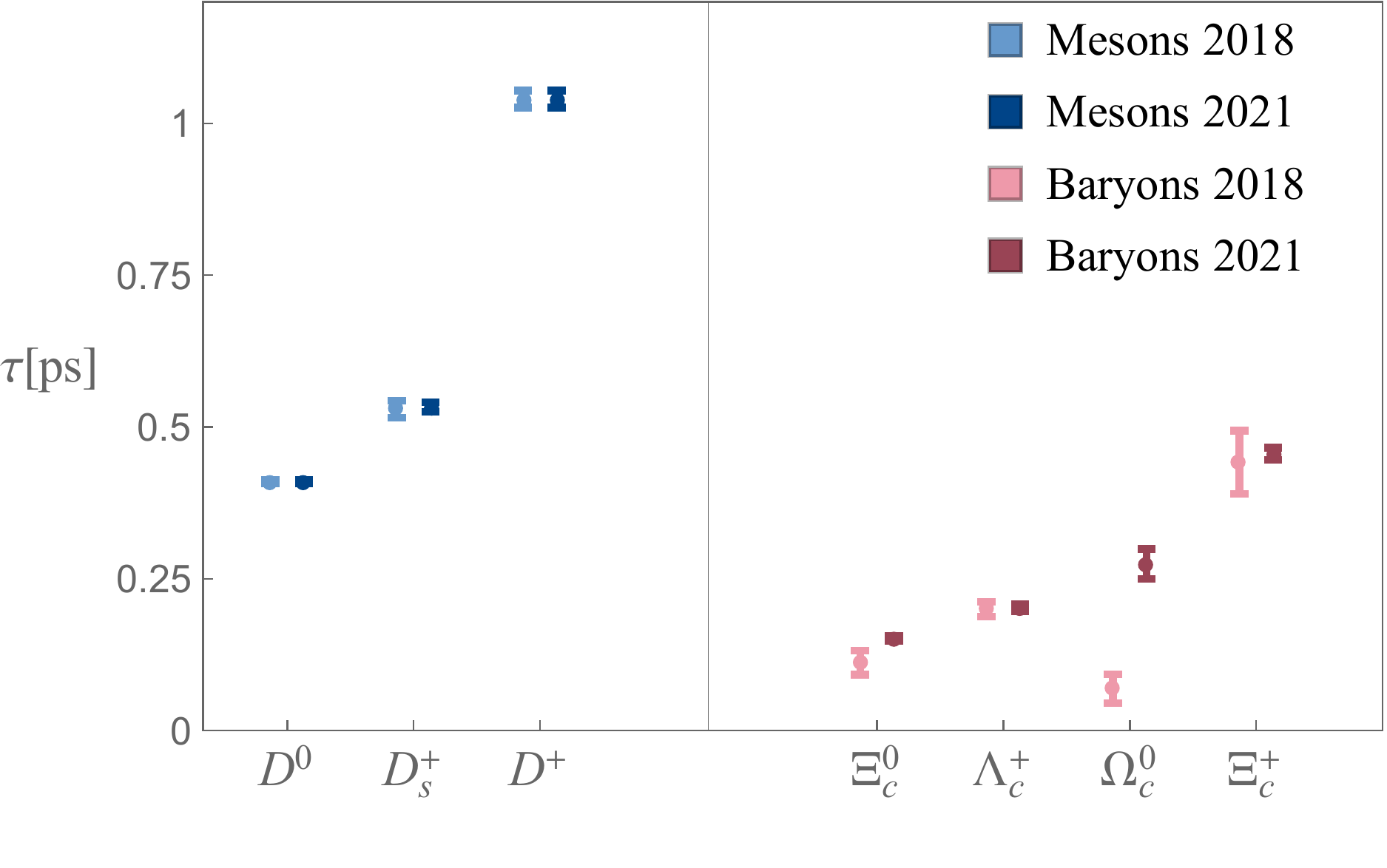}
    \caption{Changes in experimental lifetime averages for mesons (blue, on the left) and baryons (red, on the right) in 2018 (light) and 2021 (dark) owing to the recent LHCb results. Recent Belle II results \cite{BelleII2022Omegac} support the new $\Omega_c$ lifetime measurement. Note that the error bars indicate 2$\sigma$ uncertainties in all cases.}
    \label{fig:hierarchyexp}
\end{figure}

\section{Background}
\subsection{Heavy quark expansion}
Within the framework of the HQE, the decay width is expanded systematically in terms of the parameters $\Lambda_{\rm QCD}/m_c$ and $\alpha_s$; for more details, see e.g.\ \cite{Neubert1993}. It can then be presented in the, somewhat schematic, form
\begin{align}
\Gamma(H) = &~\frac{G_F^2 m_c^5}{192\pi^3} \bigg[c_3 + \frac{c_\pi\mukin+c_G\muG}{m_c^2} +  \frac{c_\rho\darwin}{m_c^3} + \dots + \frac{16 \pi^2}{2m_H} \bigg( \sum\limits_{i,q}\frac{c_{6,i}^q\langle H| \Opsix{i}{q}| H\rangle}{m_c^3} \nonumber \\
&{}+\sum\limits_{i}\dfrac{c_{7,i}^q\langle H| \OpsevenP{i}{q}| H\rangle}{m_c^4}+ \dots \bigg) \bigg]\,, 
\label{eq:DecayRateExpansion}
\end{align}
where $c_{3,\pi,G}$, etc., contain contributions from short-distance physics (Wilson coefficients, CKM factors, and dependence on non-zero quark and lepton masses), summed over all possible decay modes, while $\mukin,\, \muG,\,\darwin$, $\langle H| \Opsix{i}{q}| H\rangle$, and $\langle H| \OpsevenP{i}{q}| H\rangle$ are non-perturbative matrix elements sensitive to the decaying hadron $H$. Each of the coefficients $c_i$ also contains, implicitly, the $\alpha_s$ expansion. In fact, \eqref{eq:DecayRateExpansion} represents two separate series in $1/m_c$. The first series, consisting of two-quark operators and being a two loop effect in the OPE, can be associated with free charm quark decay, up to corrections arising in the HQE. Apart from the matrix elements $\mukin(H)$, $\muG(H)$, and $\darwin(H)$, with explicit definitions available for example in \cite{DMT2006}, this first series does not drive any significant effect on lifetime splittings, although the additionally $1/m_c$-suppressed contribution from $\darwin$, first computed for the full decay width in \cite{LPR2020,MMP2020,Moreno2020,KLPRRV2021}, turns out to be a non-negligible contribution to the overall lifetime.

\begin{figure}[ht]
    \centering
   \includegraphics[width=0.32\textwidth]{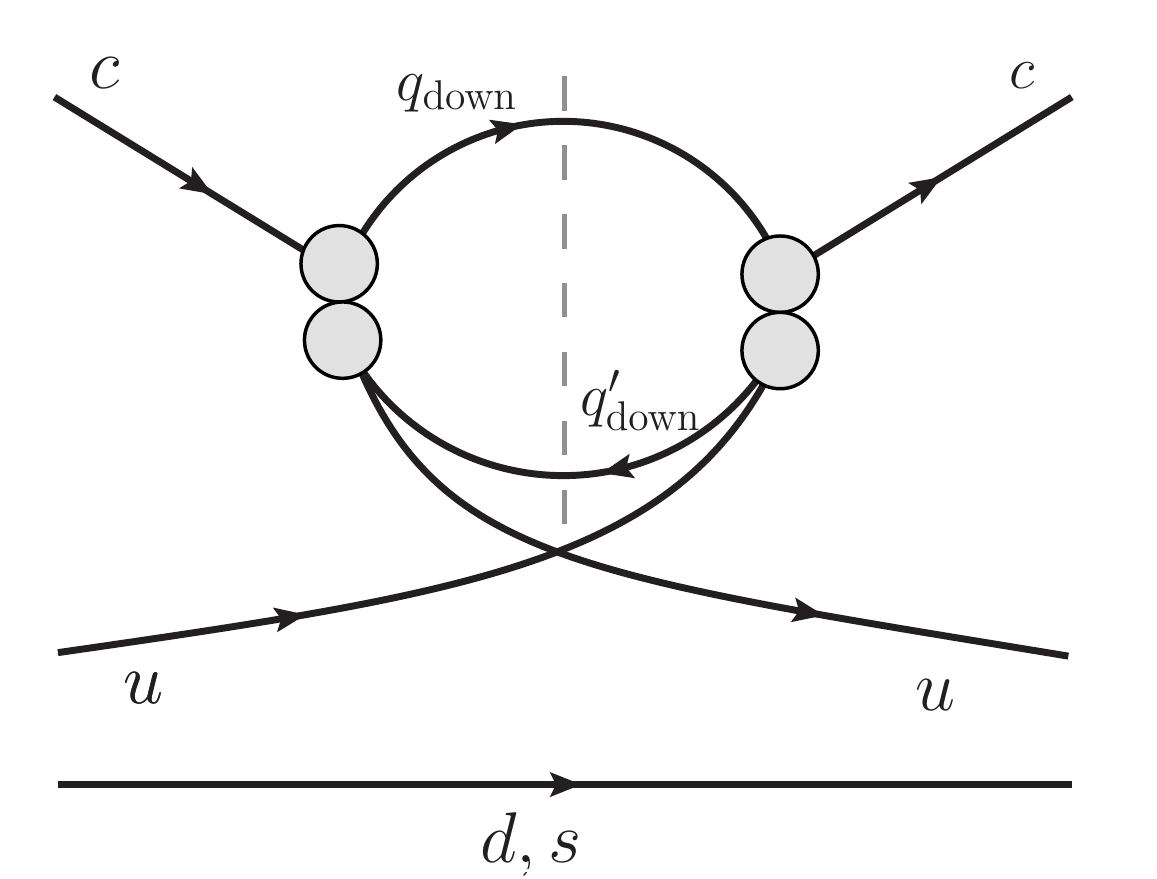}
   \includegraphics[width=0.3\textwidth]{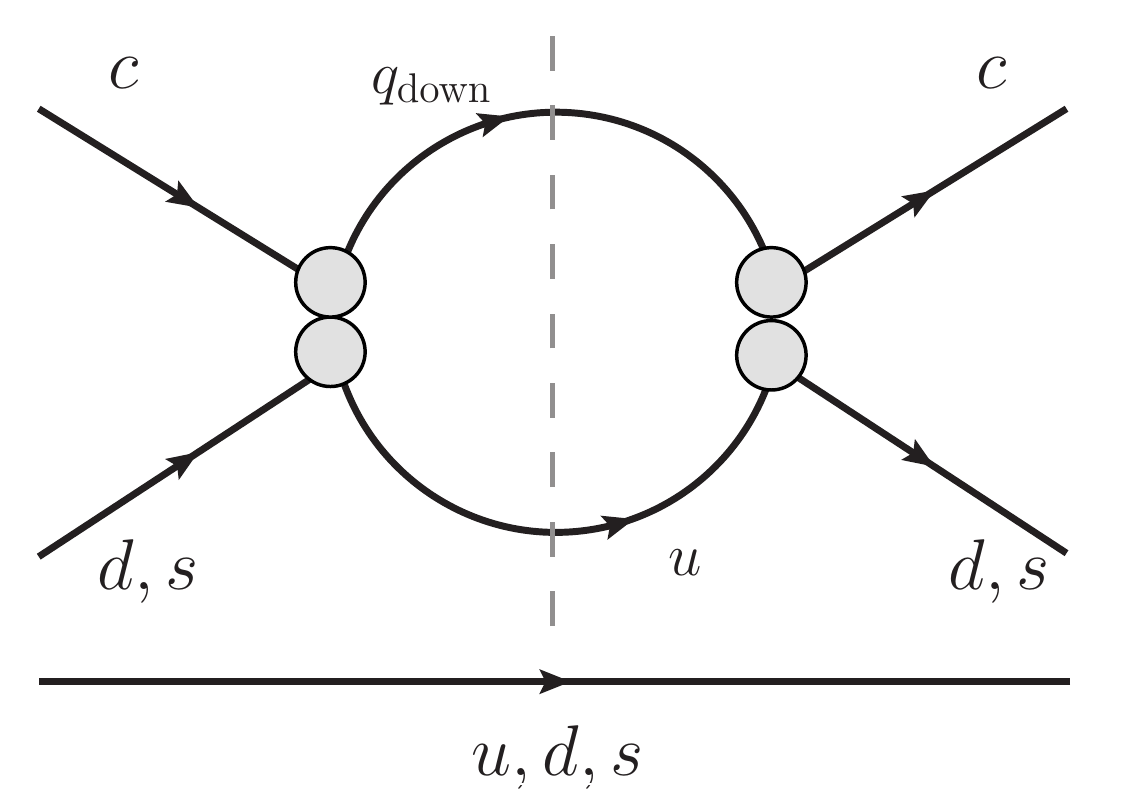}
   \includegraphics[width=0.34\textwidth]{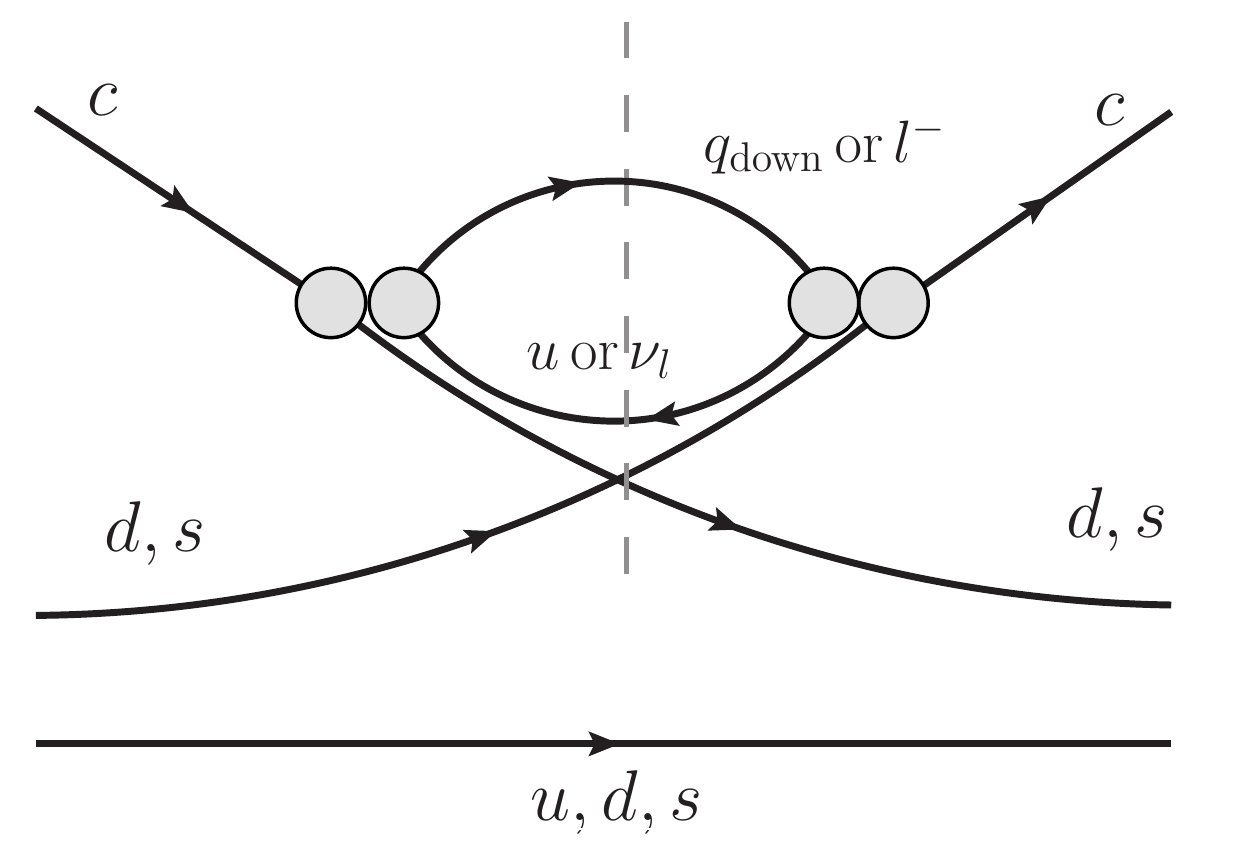}
       \caption{\small Diagrammatic representations four-quark contributions to baryon decay widths. From left to right: (a) destructive Pauli interference, labelled `$\intm$'; (b) Weak exchange, labelled `$\exc$'; (c) constructive Pauli interference, labelled `$\intp$'.}
     \label{fig:4qtopologiesBaryons}
\end{figure}

The second series, beginning with the matrix elements $\langle H| \Opsix{i}{q}| H\rangle$, represents contributions of four-quark operators, leading to the three topologies shown in figure~\ref{fig:4qtopologiesBaryons}. As such contributions are sensitive to the flavour of the light valence quarks in the hadron, and as these one-loop diagrams are additionally enhanced by the $16\pi^2$ loop factor relative to the first series, it is these terms which primarily drive the lifetime hierarchies observed in figure~\ref{fig:hierarchyexp}. Indeed, these four-quark contributions can even numerically dominate the ``leading'' contribution from free charm decay.

The number of available terms in both series has grown over time. Initially, only the leading term $c_3$ was considered. Upon the discovery of the $D^0$-$D^+$ lifetime splitting, the importance of the four-quark contributions was realised for the first time \cite{GNPR1979,KS1983,BGT1984,SV1985,SV1986}. Subsequently, the development of the HQE in the 1990s led to the introduction of the $\mukin,\, \muG$, and $\darwin$ terms, although in the case of $\darwin$ the full contribution has only recently become available for charm decays \cite{KLPRRV2021}. Attempts to explain the then-anomalous $\tau(\Lambda_b^0)/\tau(B_d^0)$ measurement prompted investigation of the higher-order four-quark contributions in the $1/m_c$ series \cite{GOP2003,GOP2004,LenzRauh2013,Cheng18c}, and $\alpha_s$ corrections in their Wilson coefficients \cite{CFLM2001,FLMT2002,LenzRauh2013}. For charmed hadrons, these contributions turn out to constitute sizeable corrections to the leading four-quark contribution. Further progress is still being made, particularly in the semileptonic contributions, with higher-order $\alpha_s$ contributions available even up to $\mathcal{O}(\alpha_s^3)$ recently \cite{FSS2020,CCD2021}; however, in our work we have kept only available LO and NLO $\alpha_s$ corrections everywhere, in order to keep consistency in expansions.

\subsection{Non-perturbative parameters}
The non-perturbative parameters, the matrix elements appearing in \eqref{eq:DecayRateExpansion}, represent the biggest challenge in the calculation of heavy hadron lifetimes. 

For mesons, these have been considered fairly extensively, with up-to-date sum rules calculations in \cite{KLR2021} for four-quark dimension-six matrix elements, and some experimental fits using inclusive semileptonic $B$ meson decays in e.g. \cite{BCG2021,Bernlochner:2022ucr} for the remaining parameters. 

For baryons, the picture is more complicated. The four-quark matrix elements are expressed in terms of the baryon wave function at the origin, and the non-relativistic quark model of de Rujula, Georgi, and Glashow \cite{GGR1975} is applied, where $|\Psi_{cq}(0)|^2$ can be extracted from the spin-spin interaction between the $c$-quark and a $qq'$-diquark in a baryon. This has been applied to baryons in several previous papers; our approach most closely follows that of Rosner \cite{Rosner1996}, in which, for example, the $\Psi_{cq}^{{\Lambda_c^+}}(0)$ wave function is expressed in terms of the $\Sigma_c^*-\Sigma$ hyperfine splittings, and normalized to a $D_q$-meson wave function:
\begin{eqnarray}
 |\Psi_{cq}^{{\Lambda_c^+}}(0)|^2 &=& y_q \frac{4}{3} \frac{M_{\Sigma_c^\ast} - M_{\Sigma_c}}{M_{D^{\ast}} - M_D} \; |\Psi_{cq}^{D_q}(0)|^2 \,, \quad |\Psi_{cq}^{D_q}(0)|^2 = \frac{1}{12} f_{D_q}^2 M_{D_q} \,,
\label{eq:RosWF}
\end{eqnarray} 
with the new ingredient of the factor $y_q$, relative to \cite{Rosner1996}, accounting for non-equal constituent quark masses in mesons and baryons \cite{KR2014}. All other baryons have similar relations, with $SU(3)_F$-breaking effects taken into account.\footnote{Note that, in \cite{GMN2022}, the operators were taken to be in the QCD basis; for a treatment in HQET, see e.g.\ \cite{GLMNPR2023}.} The dimension-seven matrix elements are then estimated by scaling relations with respect to those of dimension-six.
For the two-quark parameters, $\muG$ and $\mukin$ can be extracted from spectroscopic relations (although with more assumptions needed in the case of $\mukin$), while $\darwin$ can be expressed at leading order in $1/m_c$ in terms of the four-quark matrix elements using the equations of motion for the gluon field, giving the values in table \ref{tab:nonpertNScharm}.

\begin{table}[ht]
    \centering
    \begin{tabular}{|c|c|c|c||c|c|c|c|} \hline
        & $D^0$ & $D^+$& $D_s^+$ & $\Lambda_c^+$ & $\Xi_c^+$ & $\Xi_c^0$ & $\Omega_c^0$  \\ \hline
      $\muG/ \GeV^2$  & $0.41(12) $ & $0.41(12)$ & $0.44(13)$ & 0 & 0 & 0 & $0.26(8)$ \\ 
      $\mukin/ \GeV^2$ & $ 0.45(14) $ & $ 0.45(14) $ & $0.48(14)$ & $0.50(15)$ & $0.55(17)$ & $0.55(17)$ & $0.55(17)$ \\
      $\darwin/ \GeV^3$ &$0.056(12)$ & $0.056(22)$&$0.082(33)$ & $0.04(1)$ & $0.05(2)$ &$0.06(2)$ & $0.06(2)$ \\ \hline
    \end{tabular}
    \caption{\small Non-perturbative parameters, and their uncertainties, for all hadrons considered. For mesons, the Darwin parameters are taken from \cite{KLPRRV2021}. 
    }
    \label{tab:nonpertNScharm}
\end{table}

\subsection{Charm quark mass schemes}
The expression \eqref{eq:DecayRateExpansion} has a leading dependence on the charm quark mass as $m_c^5$, and is therefore highly sensitive to both its value and definition. While the series is traditionally expressed in terms of the pole mass, this proves to be problematic, owing to a renormalon divergence (for a review of renormalons, see \cite{Beneke1998,Beneke2021}), visible for example in the $\overline{\rm MS}$ expansion of the mass:\footnote{In the first version of \cite{Beneke2021}, there were several numerical errors in the tables with the $m_c$ and $m_b$ mass expansions, which have subsequently been corrected in the second version of the arXiv paper after we communicated them to the author.}
\begin{equation}
m_{c}^{\rm pole}= \overline{m}_c(\overline{m}_c)(1+0.16+0.15+0.21+\ldots) \,,
\label{eq:RelationOS-MS}
\end{equation}
where the series starts to increase already at the three-loop order. 
This prompts the consideration of alternative, renormalon-free mass definitions, with a schematic form, in some scheme $X$,
\begin{eqnarray}
m_{c}^X(\mu_f) &=& m_c^{\rm pole}  - \delta m_{c}^{X}(\mu_f)
\nonumber \\
&=& \overline{m}_c(\overline{m}_c) + \overline{m}_c(\overline{m}_c)    \sum_{n=1}^{\infty} \left [  c_n(\mu,\overline{m}_c(\overline{m}_c))   - \frac{\mu_f}{\overline{m}_c(\overline{m}_c) } s_{n}^X (\mu/\mu_f) \right ]  \alpha_s^n(\mu) \,.
\nonumber \\
\label{eq:mCX}
\end{eqnarray}
 One such renormalon-free mass scheme, developed for $B$ mesons, is the kinetic scheme \cite{BSUV1996}; however, for charmed hadrons this runs into problematic constraints on the factorisation scale $\mu_f$ appearing in the definition \eqref{eq:mCX}: $\Lambda_{\rm QCD} \ll \mu_f \ll m_c $, which cannot be easily satisfied. We therefore also consider the $\overline{\rm MS}$ and MSR \cite{HJS2008} schemes, where the latter can be viewed as interpolating between the pole and $\overline{\rm MS}$ schemes. At the present level of numerical accuracy, the predictions from each scheme are fully compatible with each other, but this remains an important input and will need to be considered more critically, particularly as higher-order terms in the $\alpha_s$ expansion become available in future. An alternative approach is to use a physical mass definition, e.g.\ extracting the mass from moments of $e^+e^- \to hadrons$, recently revisited in \cite{BMV2023}.

\section{Results}
In this section, we present our results for lifetimes of the lowest-lying singly charmed hadron states. Note that, for the $D_s^+$ meson, the decay $D_s^+ \to \tau^+  \nu$ is not accessible within the HQE, so we define a modified decay width $\bar\Gamma(D_s^+) = \Gamma(D_s^+)(1-BR(D_s \to \tau^+ \nu))$. Lifetime ratios and semileptonic branching ratios are defined by normalising with respect to experimental values, i.e.\
\begin{equation}
    \frac{\tau(H_1)}{\tau(H_2)} = 1 + \left(\Gamma^{\rm th}(H_2) - \Gamma^{\rm th}(H_1) \right) \tau^{\rm exp}(H_1) \,,  \quad BR^{(e)}(H_1) = \Gamma^{(e)}(H_1) \tau^{\rm exp}(H_1) \,,
\end{equation}
where $\Gamma^{(e)}(H_1) = \Gamma(H_1 \to X e \nu )$. Uncertainties arise from variations in the hadronic and scale parameters, with an additional 30\% uncertainty assigned to the baryon wave function to account for model dependence; and from variations in the renormalisation scale. The resulting uncertainty estimates are fairly conservative, but there is some cancellation in the ratios defined above.

\subsection{Mesons}

\begin{table}[ht]
\small
\renewcommand{\arraystretch}{1.4}
\centering
\begin{tabular}{|c|c|c|c|c||c|}
\hline
Observable & Pole & $\overline{\text{MS}}$ &  Kinetic & MSR & Experiment \\
\hhline{|=|=|=|=|=||=|}
$\Gamma(D^0)$  & $1.71^{+0.41+0.39}_{-0.47-0.36}$ & $1.43^{+0.36+0.48}_{-0.40-0.40}$  &  $1.77^{+0.40+0.53}_{-0.45-0.45}$  & $1.68^{+0.38+0.53}_{-0.43-0.44}$ & $2.44\pm 0.01$\\
\hline
$\Gamma(D^+)$ & $-0.07^{+0.76+0.31}_{-0.68-0.20}$ & $-0.27^{+0.66+0.03}_{-0.88-0.04}$ & $-0.07^{+0.73+0.20}_{-0.66-0.14}$ & $-0.13^{+0.71+0.13}_{-0.64-0.11}$ & $0.96\pm 0.01$\\
\hline
$\tilde{\Gamma}(D_s^+)$ & $1.71^{+0.49+0.44}_{-0.60-0.40}$ & $1.43^{+0.42+0.49}_{-0.52-0.41}$ & $1.77^{+0.47+0.55}_{-0.58-0.47}$ & $1.67^{+0.46+0.55}_{-0.56-0.46}$ & $1.88\pm 0.02$\\
\hhline{|=|=|=|=|=||=|}
$\tau(D^+)/\tau(D^0)$ & $2.85^{+0.68+0.10}_{-0.81-0.17}$ & $2.78^{+0.63+0.47}_{-0.73-0.37}$ &$2.91^{+0.68+0.35}_{-0.80-0.32}$ & $2.89^{+0.66+0.42}_{-0.78-0.35}$ & $2.54\pm 0.02$\\
\hline
$\tilde{\tau}(D_s^+)/\tau(D^0)$ & $1.00^{+0.24+0.02}_{-0.22-0.02}$ &
$1.00^{+0.21+0.01}_{-0.19-0.00}$&
$1.00^{+0.23+0.01}_{-0.21-0.01}$ & $1.00^{+0.23+0.01}_{-0.21-0.01}$&  $1.30\pm 0.01$\\
\hline
\end{tabular}
\caption{\small Total decay widths in units $\text{ps}^{-1}$, and their ratios for charmed mesons, compared to the experimental values \cite{PDG2022}, in various mass schemes as defined in \cite{GMN2022}. Uncertainties arise from parametric (first) and renormalisation scale (second) variations.}
\label{tab:fin_mesons}
\end{table}

\begin{table}[ht]
\small
\renewcommand{\arraystretch}{1.4}
\centering
\begin{tabular}{|c|c|c|c|c||c|}
\hline
Observable & Pole & $\overline{\text{MS}}$ &  Kinetic & MSR & Experiment \\
\hhline{|=|=|=|=|=||=|}
$BR^{(e)}(D^0)\,[\%]$ & $4.07^{+2.21+0.84}_{-2.53-0.97}$ & $5.18^{+1.59+0.63}_{-1.82-0.55}$ & $5.87^{+1.94+0.22}_{-2.23-0.19}$ & $5.86^{+1.80+0.48}_{-2.07-0.41}$ & $6.49\pm 0.16$\\
\hline
$BR^{(e)}(D^+)\,[\%]$  &
$10.34^{+5.69+2.12}_{-6.52-2.44}$&
$13.15^{+4.10+1.61}_{-4.73-1.40}$ & $14.92^{+5.00+0.57}_{-5.75-0.49}$& $14.90^{+4.67+1.22}_{-5.37-1.06}$& $16.07\pm 0.30$\\
\hline
$BR^{(e)}(D_s^+)\,[\%]$ & $5.42^{+3.02+0.96}_{-3.44-1.10}$ &
$6.86^{+2.42+0.83}_{-2.83-0.72}$&
$7.67^{+2.80+0.34}_{-3.23-0.29}$ & $7.67^{+2.67+0.65}_{-3.10-0.56}$&  $6.30\pm 0.16$\\
\hhline{|=|=|=|=|=||=|}
$\Gamma^{(e)}(D^+)/\Gamma^{(e)}(D^0)$ & $1.00^{+0.02+0.00}_{-0.02-0.00}$ &
$1.00^{+0.01+0.00}_{-0.01-0.00}$&
$1.00^{+0.02+0.00}_{-0.02-0.00}$ & $1.00^{+0.02+0.00}_{-0.01-0.00}$&  $0.977\pm 0.031$\\
\hline
$\Gamma^{(e)}(D_s^+)/\Gamma^{(e)}(D^0)$ & $1.05^{+0.29+0.01}_{-0.31-0.01}$ &
$1.06^{+0.24+0.01}_{-0.27-0.01}$&
$1.07^{+0.28+0.01}_{-0.30-0.01}$ & $1.06^{+0.26+0.01}_{-0.29-0.01}$&  $0.790\pm 0.026$\\
\hline
\end{tabular}
\caption{\small Semileptonic decay widths in inclusive channel $D_{(s)} \to X e\nu$ in units $\text{ps}^{-1}$, and their ratios for charmed mesons compared to the experimental values \cite{PDG2022}, in various mass schemes as defined in \cite{GMN2022}.}
\label{tab:fin_mesonsSL}
\end{table}

Our results for meson inclusive and and semileptonic observables, for various mass schemes, are presented in tables \ref{tab:fin_mesons} and \ref{tab:fin_mesonsSL} respectively. Compared with experimental values, we observe some slight tensions, most notably in the prediction for $\Gamma(D^+)$, which is found to be negative (i.e. unphysical) for central values of the input parameters. This is due to a significant negative contribution from Pauli interference. However, the ratio $\tau(D^+)/\tau(D^0)$ is found to be compatible with experiment. On the other hand, $\bar\tau(D_s^+)/\tau(D^0)$ is predicted to be closer to unity than is observed in experiment, while the individual decay widths are both compatible with the experimental values. These observations have also been made in other recent studies of inclusive charmed mesons \cite{KLPRRV2021,Cheng18c}. The picture could be improved with additional contributions to the inclusive decay widths, and with better control of the nonperturbative parameters, as the theoretical uncertainties are quite large, even in the ratios.

\subsection{Baryons}

\begin{figure}[htb]
    \centering
    \includegraphics[width=0.8\textwidth]{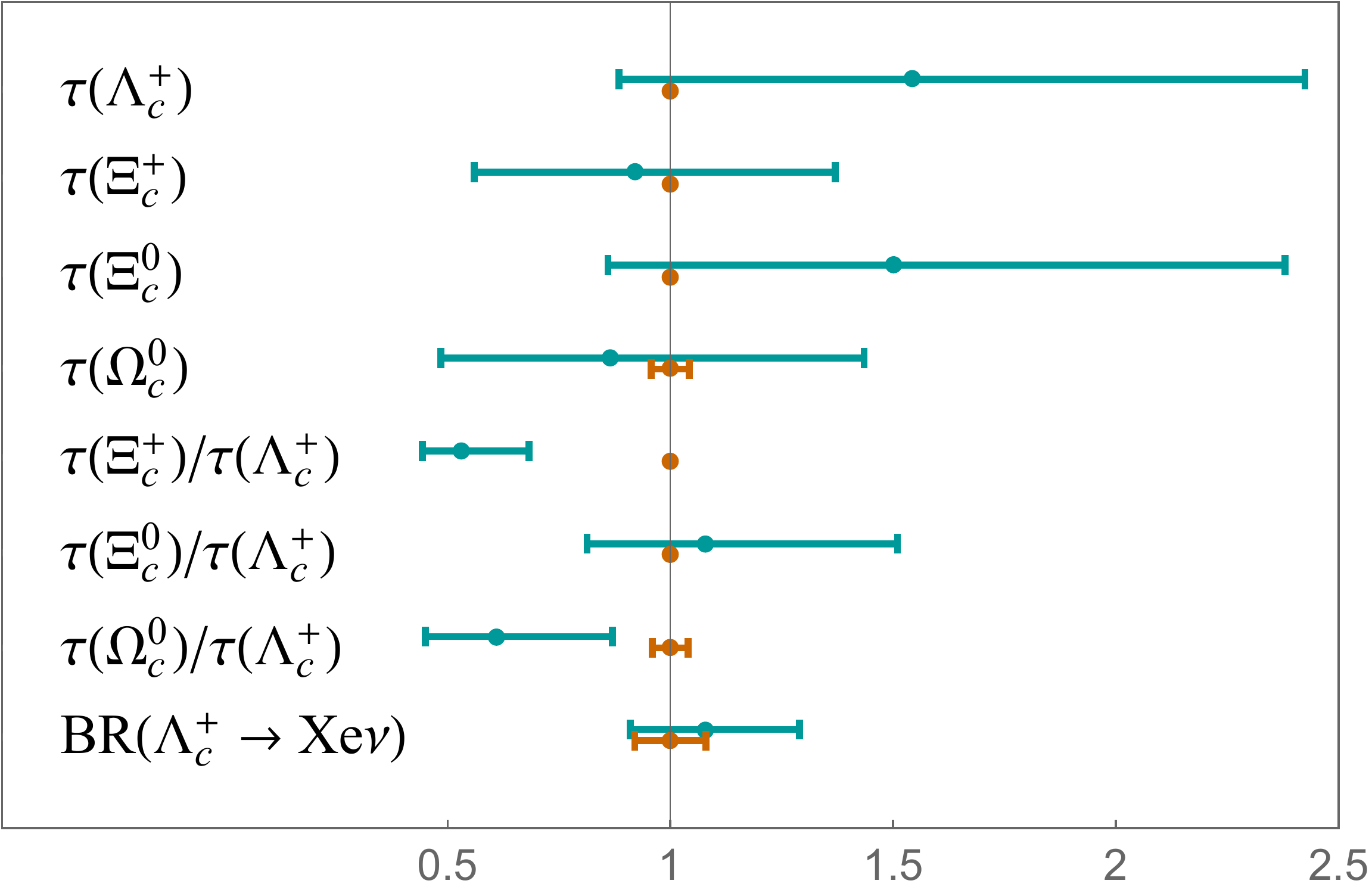}
    \caption{Predictions for observables of charmed baryons, normalised to corresponding experimental values, in the MSR scheme. Experimental values, in orange, also have uncertainties indicated when they are larger than 1\%. }
    \label{fig:results_baryons_MSR}
\end{figure}
\begin{figure}[htb]
    \centering
\includegraphics[width=0.9\textwidth]{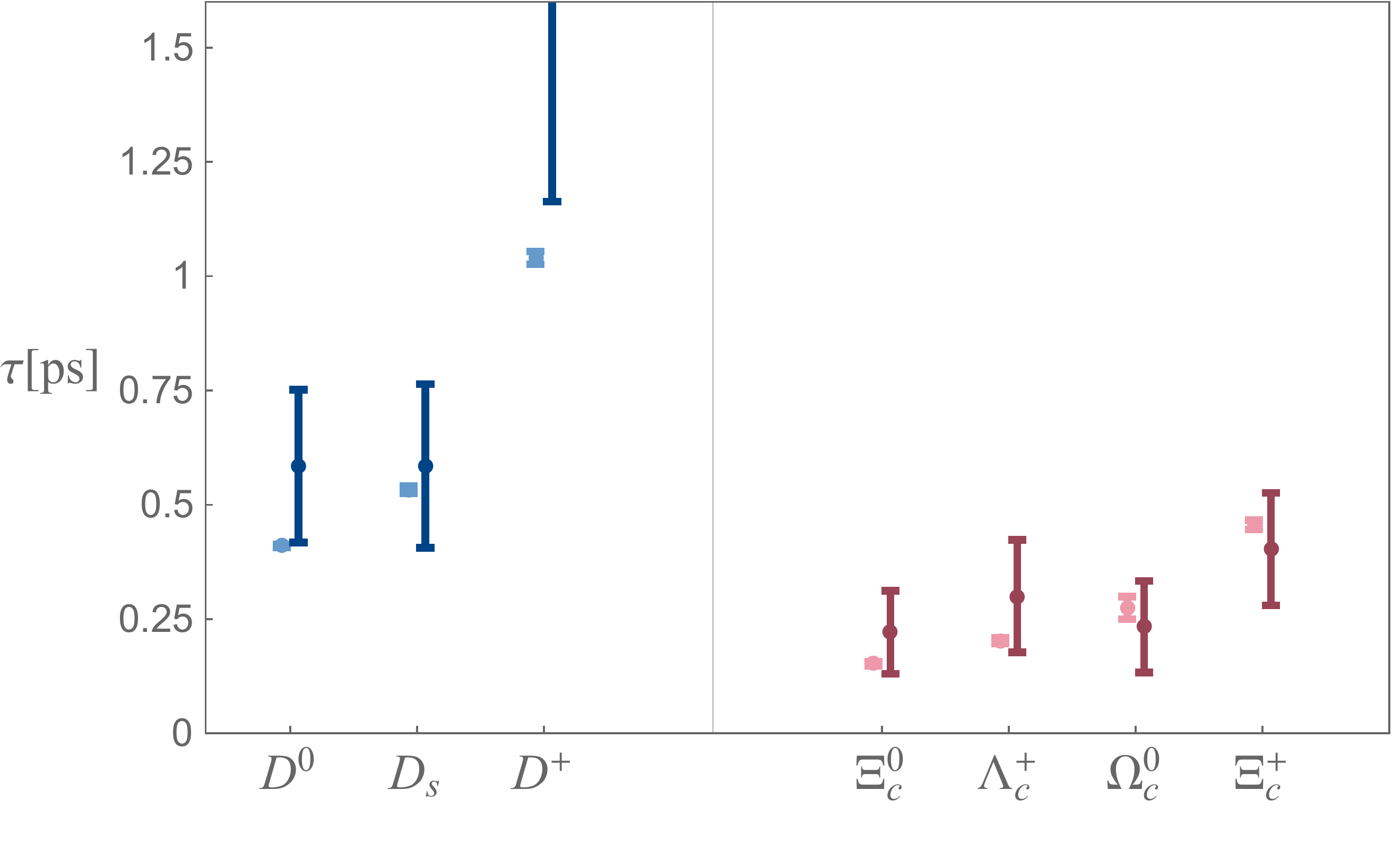}
\caption{\small Hierarchy of lifetimes of charmed mesons (left, in blue) and singly charmed baryons (right, in red). Our predictions, in the kinetic scheme, are compared to the latest experimental values (left of each pair of values) \cite{LHCbOmegac2021,PDG2022}.}
\label{fig:hierarchyexpth}
\end{figure}

Results for observables for singly charmed baryons are presented in figure~\ref{fig:results_baryons_MSR}, for the MSR mass scheme, where they are compared to experimental predictions. As can be seen, the results are broadly compatible with experiment. In particular, we find that our prediction for the $\Omega_c^0$ lifetime is consistent with the new data \cite{LHCbOmegac2018,LHCbcharmedLifetimes2019,LHCbOmegac2021,BelleII2022Omegac}, and contradicts an earlier conjecture in \cite{Cheng18c} that the HQE might fail for the $\Omega_c^0$. Some tensions again arise, though: we predict lower values for the two ratios $\tau(\Xi_c^+)/\tau(\Lambda_c^+)$ and $\tau(\Omega_c^0)/\tau(\Lambda_c^+)$ than seen in experimental measurements, a conclusion that may be attributable to our overestimate of the central value of $\tau(\Lambda_c^+)$. Note that, among the baryons, only the $\Lambda_c^+$ semileptonic branching ratio has been measured, which prevents a complete comparison of all observables. 
However, our predicted values for semileptonic branching ratios, which significantly differ between the baryons and are presented in table \ref{tab:BaryonSL}, are important for assessing the validity of HQE in charmed baryons, and so experimental measurements of $BR(\Xi_c^+ \to X e \nu)$, $BR(\Xi_c^0 \to X e \nu)$, and $BR(\Omega_c^0 \to X e \nu)$ are needed.
\begin{table}[th]
\renewcommand{\arraystretch}{1.4}
\centering
\begin{tabular}{|c|c|}
\hline
$BR(\Lambda_c^+ \to X e \nu)$/\% & $4.28^{+0.47+0.39}_{-0.37-0.30}$\\
\hline
$BR(\Xi_c^+ \to X e \nu)$/\% &  $14.95^{+2.66+1.59}_{-2.45-1.50}$ \\
\hline
$BR(\Xi_c^0 \to X e \nu)$/\%  & $5.06^{+0.91+0.54}_{-0.84-0.51}$\\
\hline
$BR(\Omega_c^0 \to X e \nu)$/\% & $11.19^{+3.01+1.94}_{-2.89-2.09}$\\
\hline
\end{tabular}
\caption{\small Results for semileptonic branching fractions in the MSR mass scheme. The first and second errors correspond to hadronic and renormalization scale uncertainties, respectively.}
\label{tab:BaryonSL}
\end{table}
Again, the relatively slow convergence of the $1/m_c$ expansion suggests that further contributions might alter the picture, but in general we see broad agreement between our predictions and experiment across all charmed hadrons. This conclusion remains true across all mass schemes considered in the paper, which we find to be indistinguishable at the present level of numerical accuracy. In figure~\ref{fig:hierarchyexpth}, we compare our individual lifetime predictions to the latest experimental hierarchy.

\section{Conclusions}
We have presented the most up-to-date predictions for lifetimes, and lifetime ratios, of singly-charmed hadrons. In particular, we apply new results for the Darwin term to the baryon sector. Our results show that the newly-established experimental hierarchy,
\begin{eqnarray}
\tau(\Xi_c^0) < \tau(\Lambda_c^+)< \tau(\Omega_c^0) < \tau(\Xi_c^+)\,,
\end{eqnarray}
is consistent with the theory prediction, albeit with large uncertainties on the theory side. Some tensions do exist, however, most notably in  $\Gamma(D^+)$, while the ratios $\tau(\Xi_c^+)/\tau(\Lambda_c^+)$ and $\tau(\Omega_c^0)/\tau(\Lambda_c^+)$ are in tension with experiment. Since the predicted individual lifetimes are in agreement with experiment, within uncertainties, the  tension in the lifetime ratios can be attributed to an overly high central value of the $\Lambda_c^+$ lifetime prediction.  Our results favour the new measurements of $\tau(\Omega_c^0)$, which strongly suggests that the HQE remains applicable to charm decays, and contradicts the suggestion in \cite{Cheng18c} that the HQE fails specifically for the $\Omega_c^0$.

In light of this, further work increasing the number of available terms in the $1/m_c$ and $\alpha_s$ expansions, along with better control of the input parameters and their uncertainties, such as via a lattice computation, would be beneficial. Alongside this, the questions as to how best to formulate the HQE for the charm quark, and what is the optimal mass scheme for $m_c$ in order to control divergent behaviour in the $\alpha_s$ expansion, remain important \cite{FMV2019,MMP2021}, as does an exploration of quark-hadron duality in charm processes (recently analysed in the context of heavy meson decays in \cite{Umeeda2021}).

Combined with the works in \cite{KLPRRV2021,GLMNPR2023,LPR2022}, this constitutes the most up-to-date predictions for lifetimes of all the lowest-lying heavy hadrons containing a single charm or bottom quark.

\subsection*{Acknowledgments}
\noindent
B.M. thanks organizers for their effort and the very pleasant atmosphere they created at the workshop. Support of the Croatian Science Foundation (HRZZ) project, ``Heavy hadron decays and lifetimes'' IP-2019-04-7094, as well as sponsorship from the Alexander von Humboldt Foundation in the framework of the Research Group Linkage Programme, with funding from the German Federal Ministry of Education and Research, is gratefully acknowledged.

\end{document}